\documentclass[12pt,onecolumn]{IEEEtran}
\usepackage{times,amsmath,amssymb,amsfonts}

\usepackage{graphicx,subfigure}

\usepackage[usenames,dvipsnames]{pstricks}
\usepackage{epsfig}
\usepackage{pst-grad} % For gradients
\usepackage{pst-plot} % For axes
\usepackage{graphicx,subfigure}
\graphicspath{{figures/}}

\usepackage{multirow}

\DeclareMathOperator{\D}{d\!}
\DeclareMathOperator{\sfT}{\mathsf T} 
\DeclareMathOperator{\sfH}{\mathsf H}
\newcommand{\nn}{\nonumber \\}
\newcommand{\mb}{\mathbf}
\DeclareMathOperator{\diag}{diag}

\title{Optimized L2-Orthogonal STC CPM for 3 Antennas}
\author{%
Matthias Hesse\mbox{$^*$}, Jerome Lebrun and Luc Deneire%
\vspace{1mm}\\
\fontsize{10}{10}\selectfont\itshape
Laboratoire I3S, CNRS, University of Nice, Sophia Antipolis, France\vspace{1mm}\\
\fontsize{9}{9}\selectfont\ttfamily\upshape
\{hesse,lebrun,deneire\}@i3s.unice.fr\\ \vspace{1em} \texttt{\small- final version - 22.Oct.2008 -}}

\begin{document}
\maketitle

\begin{abstract} 
  ~In this paper, we introduce further our recently designed family of
  L$^2$ orthogonal Space-Time codes for CPM. With their advantage of
  maintaining both the constant envelope properties of CPM, the
  diversity of Space-Time codes and moreover orthogonality, and thus
  reduced decoding complexity, these codes are also full rate, even
  for more than two transmitting antennas.

  The issue of power efficiency for these codes is first dealt with
  by proving that the inherent increase in bandwidth in these systems
  is quite moderate.  It is then detailed how the initial state of the
  code influences the coding gain and has to be optimized.  For the
  two and three antennas case, we determine the optimal values by
  computer simulations and show how the coding gain and therewith the
  bit error performance are significantly improved by this
  optimization.  \footnote{$^*$The work of M. Hesse is supported by a
    EU Marie-Curie Fellowship (EST-SIGNAL program) under contract No
    MEST-CT-2005-021175.}
  % Space-time codes (STC) for continuous phase modulation (CPM) are
  % considered. Therefore, two different $L^2$-orthogonal STC are used
  % which were recently proposed. The advantage of these codes is the
  % reduced decoding complexity while maintaining all favorable
  % properties of CPM. Further, $L^2$-orthogonal code design is not
  % limited to two transmitting antennas but is capable to preserve
  % full rate.

  % The previously proposed design did not consider the initial state
  % of the code as a parameter. This paper shows the importance of
  % carefully chosen initial states. It is analytically shown that the
  % coding gain depends on the initial states and optimal values for
  % two and three antennas are determined by computer simulations.
  % Finally, we show that the coding gain and therewith the bit error
  % performance is significantly improved by optimization.

\end{abstract}

\section{Introduction}

The combination of space-time coding (STC) and continuous phase
modulation (CPM) is an attractive field of research because the
combination of STC and CPM brings the benefits of diversity and low
power consumption to wireless communications.  Zhang and Fitz
\cite{Zhan00} were the first to apply this idea by constructing a
trellis based scheme. This idea was pursued by Zaji\'c and St\"uber in
\cite{Zaji05} for full response CPM, afterward optimized \cite{Zaji06}
and extended to partial response CPM in \cite{Zaji07}. But for these
codes the decoding effort grows exponentially with the number of
transmitting antennas.  This was circumvented by burst-wise
orthogonality introduced by Silvester, Schober and Lampe in
\cite{Silv05} and by block-wise orthogonality established by Wang and
Xia \cite{Wang04}\cite{Wang05}.  Unfortunately, based on the Alamouti code
\cite{Alam98}, this latter design is restricted to two antennas. This
restriction was partially circumvented in \cite{Wang03} by using quasi
orthogonal STC for 4 transmitting antennas. However, to our knowledge,
there was so far no consistent orthogonal design for a general number of
antennas.

In our previous publications, by relaxing the orthogonality condition,
we have been able to construct new $L^2$-orthogonal space-time codes
which achieve full rate and full diversity with low decoding effort.
More precisely, in \cite{Hess08a} we have generalized the two-antenna
code proposed by Wang and Xia \cite{Wang05} from pointwise to
$L^2$-orthogonality and in \cite{Hess08} and \cite{Hess08g}, we
introduce the first $L^2$-orthogonal code for CPM with three antennas.
In this paper we shortly display some of these results and focus on
several properties of these codes. Of special interest is the
optimization of the bit error rate which depends on the initial phase
of the system. Our simulation results illustrate the systemic behavior
of these conditions.

\section{Parallel Coded CPM}

\subsection{Space-Time code block}

In MIMO systems, the signal sent $s(t)$ is transmitted via $L_t$
transmitting antennas. Here, similarly to \cite{Wang05}, the
continuous phase modulated sending signal will be split into blocks

\begin{equation}
  \mathbf S(t)=\left[
    \begin{smallmatrix}
      s_{1,1}(t)&\ldots&s_{1,L_t}(t)\\
      \vdots &s_{m,r}(t)&\vdots\\
      s_{L_t,1}(t)&\ldots&s_{L_t, L_t}(t)\\
    \end{smallmatrix}\right]
  \label{eq:smatrix}
\end{equation}
where each element of the sending matrix
\begin{equation}
  s_{m,r}(t)=\sqrt{\frac{E_s}{L_tT}}\exp\left( j2\pi\phi_{m,r}(t)\right)
  \label{eq:defs}
\end{equation}
is defined within the time interval $(L_tl+r-1)T\leq t\leq (L_tl+r)T$.
The indexes $m$ and $r$ denote respectively the sending antenna and
the time slot in a block. $E_s$ is the symbol energy per transmit
antenna Tx and per symbol time $T$. The phase signal
\begin{equation}
  \phi_{m,r}(t)=\theta_m(L_tl+r)+h\sum_{i=1}^{\gamma}d_{m,r}^{(l,i)}q(t-i'T)+c_{m,r}(t)
  \label{eq:defphi}
\end{equation}
is essentially the phase of a conventional CPM signal \cite{Ande86}
with an additional correction factor $c_{m,r}(t)$. The data symbols
$d_{m,r}^{(l,i)}$ are uniquely determined for each element of the code
block. A central point of the coding scheme is the mapping between
these symbols and the data sequence $d_j$ with
$d_j\in\Omega_{d}=\{-M+1,-M+3,\ldots,M-1\}$. As in conventional CPM,
$h=2m_0/p$ is the modulation index where $m_0$ and $p$ are relative
primes. The phase smoothing function $q(t)$ is a continuous function
with $q(t)=0$ for $t\leq0$ and $q(t)=1/2$ for $t\geq\gamma T$ and
$\gamma$ is the memory length. The phase memory will be obtained from
the definitions of correction factor and the mapping of the data
symbols.

\subsection{$L^2-orthogonality$}
\label{sec:ortho}

% \begin{figure}
%   \input{mapping.tex}
%   \caption{Mapping of the data sequence to the data symbols }
%   \label{fig:mapping}
% \end{figure}

To achieve simplified blockwise decoding as in systems based on
orthogonal space-time block codes (OSTBC) \cite{Alam98}\cite{Wang05}, it is
sufficient \cite{Hess08} to force each block to be $L^2$-orthogonal
\begin{equation}
  \int\limits_{lL_tT}^{(l+1)L_tT}\mathbf S(t)\mathbf S^{\sfH}(t)\D t=E_s\mathbf I
  \label{eq:ortho}
\end{equation}
where $\mathbf I$ is the identity matrix. Hence the correlation
between two different Tx antennas $s_{m,r}(t)$ and $s_{m',r}(t)$ is
zeroed over the $l^{\mbox{th}}$ complete STC block if 
\begin{equation}
\int\limits_{lL_tT}^{(l+1)L_tT}s_{m,r}(t)s_{m',r}^*(t)\D t = 0
\end{equation}
with $m\neq m'$. Now, from theses conditions, $L^2$-orthogonal codes can be
constructed \cite{Hess08a}\cite{Hess08}\cite{Hess08g} by expressing our design
criteria:
\begin{enumerate}
\item the mapping of the data symbols $d_{m,r}^{(l,i)}$ to the data
  sequence $d_j$ and
\item the correction factor $c_{m,r}(t)$.
\end{enumerate}

\begin{figure}
  \centering
  \includegraphics[width=7cm]{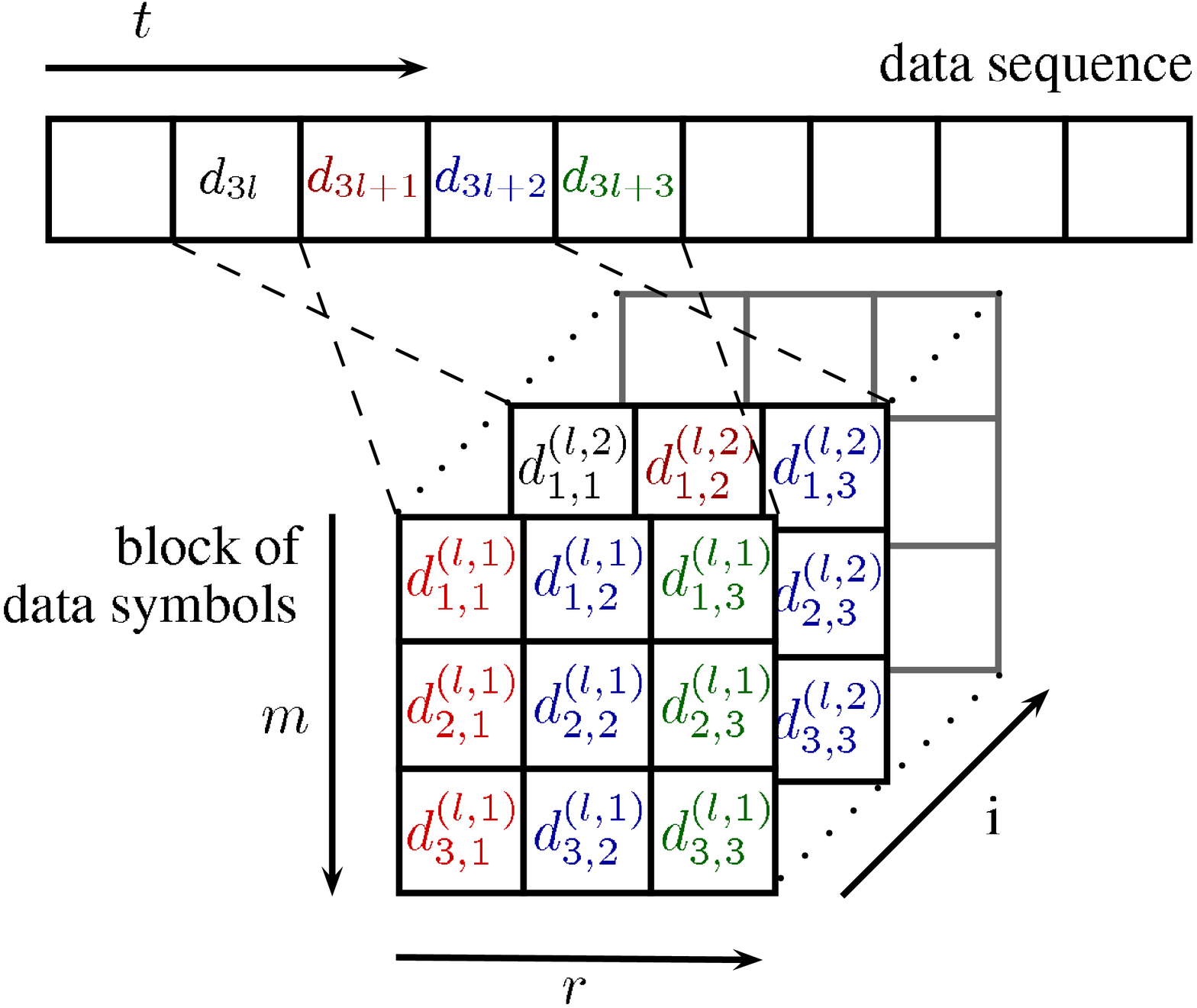}
  \caption{Mapping of the uncoded data sequence to the code block for
    3 Tx antennas}
  \label{fig:mapping}
\end{figure}

\textit{Mapping}: The most convenient mapping \cite{Hess08} is the
\textit{parallel mapping} where each data symbol of one time slot $r$
for $m=1,\ldots,L_T$ is mapped to the same symbol of the data sequence
$d_j$. The following $L_t$ data symbols at $r+1$ are mapped to
$d_{j+1}$ and so on. Each row of the code is therewith modulated by
the same symbols of the data sequence and
\begin{equation}
  d_{m,r}^{(l,i)}=d_{L_tl+r-i+1}
\end{equation}
Due to this parallel structure, this solution is named parallel codes
(PC). This mapping is illustrated in Figure \ref{fig:mapping}.

\textit{Correction factor}:
% While the mapping is similar for the two and three transmitting
% antenna code, the correction factor $c_{m,r}(t)$ has to be adapted
% to the number of antennas. For two transmitting antennas the
% correction factor has to ensure a phase shift of $\Delta\phi=1/2$
% between the first and the second antenna at $t=(2l+1)T$. This is
% achieved by setting the correction factor of the first antenna to
% zero $c_{1,r}(t)=0$. For the factor of the second transmitting
% antenna, a linear function can be used (linPC)
% \begin{equation}
%   c_{2,r}(t)=\frac{t-(2l+r)T}{2T}.
% \end{equation}
% A second possibility for the correction factor of the second antenna
% is
% \begin{equation}
%   c_{2,r}(t)=\sum\limits_{i=1}^\gamma q(t-i'T).
%   \label{eq:cor22}
% \end{equation}
% Using Eq. (\ref{eq:cor22}) with Eq. (\ref{eq:defphi}) one can see
% that the sums from each equation can be merged. We combine the
% correction factor with the data symbol and obtain a new alphabet
% $\Omega_{d_2}=\{-M+1-\frac{1}{h},
% -M+3-\frac{1}{h},\ldots,M-1-\frac{1}{h}\}$ with an offset (offPC)
% for the second antenna. This intuitive representation greatly
% simplifies modulation and demodulation.
% % \begin{align}
% %   \Omega_{d_2}=\left\{
% %     -M+1-\frac{1}{h},-M+3-\frac{1}{h},\ldots,M-1-\frac{1}{h}\right\}\nonumber.
% % \end{align}
The correction factor is determined by the phase difference
$\Delta\phi$ between the different transmitting antennas at the end of
each symbol in the code block. To ensure  phase continuity,
this difference has to be
\begin{itemize}
\item $1/2$ at $t=(2l+1)T$ for 2 Tx antennas
\item and $1/3$ or $2/3$ at $t=(3l+1)T$ and $2/3$ or $4/3$ at
  $t=(3l+2)T$ for 3 Tx antennas.
\end{itemize}
We define the linear correction factor (linPC) as
\begin{equation}
  c_{m,r}(t)=\frac{m-1}{L_t}\cdot\frac{t-(L_tl+r)T}{T}
  \label{eq:c1}
\end{equation}
and a more complex one based on the phase smoothing function $q(t)$ as
\begin{equation}
  c_{m,r}(t)=\frac{m-1}{L_t}\sum\limits_{i=1}^\gamma 2q(t-i'T).
  \label{eq:c2}
\end{equation}
Without loss of generality, the correction factor of the first antenna
is set to $c_{1,r}(t)=0$. Using Eq. (\ref{eq:c2}) with Eq.
(\ref{eq:defphi}) one can see that the sums from each equation can be
merged. So, by combining the correction factor with the data symbol,
we get new alphabets for each transmitting antenna $m$:
\begin{displaymath}
  \textstyle  \Omega_{d_m}=\{{\scriptstyle -M+1+\frac{2(m-1)}{L_th},-M+3+\frac{2(m-1)}{L_th},\ldots,M-1+\frac{2(m-1)}{L_th}}\}.
\end{displaymath}
Consequently, this code is called offset PC (offPC) and may be seen as
$L_t$ conventional CPM signals with different alphabet sets
$\Omega_{d_m}$ for each antenna $m$. The alphabet of the first antenna
$\Omega_{d_1}$ is equal to the alphabet of a conventional CPM
$\Omega_d$. For example, for two transmitting antennas we have
$c_{1,r}(t)=0$, $c_{2,r}=\sum_{i=1}^\gamma q(t-i'T)/2$ and then
$\Omega_{d_1}=\Omega_{d}=\{-M+1,-M+3,\ldots,M-1\}$ and
$\Omega_{d_2}=\{-M+1+1/h,-M+3+1/h,\ldots,M-1+1/h\}$. This intuitive
representation greatly simplifies modulation and demodulation
\cite{Hess08}.
% \begin{align}
%   &\Omega_{d_2}=\left\{
%     -M+1+\frac{2}{3h},-M+3+\frac{2}{3h},\ldots,M-1+\frac{2}{3h}\right\}\nn
%   &\Omega_{d_3}=\left\{
%     -M+1-\frac{2}{3h},-M+3-\frac{2}{3h},\ldots,M-1-\frac{2}{3h}\right\}\nonumber.
% \end{align}

With these definitions of the coding scheme, we can rewrite the
correction factor in the more generic form
\begin{equation}
  \theta_m(L_tl+r+1)=\theta_m(L_tl+r)+\xi(L_tl+r)
\end{equation}
where the function $\xi(L_tl+r)$ guarantees the continuity of the
phase for any correction factor and mapping. For parallel mapping
(similarly to conventional CPM) and $c_{m,r}(t)=0$, we get
$\xi(L_tl+r)=\frac{h}{2}d_{L_tl+1-\gamma}$.

% \subsection{Matrix representation}
% 
% For convenience, CPM code blocks $\mb S(t)$ can be decomposed into a
% matrix product:
% \begin{equation}
%   \mathbf S(t)=\mb\Theta\cdot\mb C(t)\cdot\mb\Xi\cdot\mb D(t)
% \end{equation}
% where $\mb\Theta$, $\mb C(t)$ and $\mb D(t)$ are $L_t\times L_t$
% diagonal matrices with the inter-block phase memory matrix
% \begin{equation}
%   \mathbf{\Theta}=\diag[\exp\big(j2\pi\theta_m(3l+1)\big)],
% \end{equation}
% the correction matrix
% \begin{equation}
%   \mathbf C(t)=\diag[\exp\big(j2\pi c_m(t)\big)]
% \end{equation}
% and the data matrix
% \begin{equation}
%   \mathbf D(t)=\diag[\exp\big(j2\pi h\sum\limits_{i=1}^\gamma d_{3l+r-i+1}q(t-i'T)\big)]
% \end{equation}
% for $m=1\ldots L_t$. The inner-block phase memory matrix $\mb \Xi$
% is a $L_t\times L_t$ square matrix
% 
% \begin{equation}
%   \mathbf \Xi=\left[\begin{smallmatrix}
%       1& \exp\big(j2\pi \xi_1(3l+1)\big) &\cdots&\exp\big(j2\pi \xi_1(3l+Lt-1)\big)\\
%       \vdots&\vdots&\vdots&\vdots\\
%       1& \cdots %\exp\big(j2\pi \xi_{L_t}(3l+1)\big) 
%       &\cdots&\exp\big(j2\pi \xi_{L_t}(3l+L_t-1)\big)
%     \end{smallmatrix}\right].
% \end{equation}

\section{Initial Phase}

Our model includes now all the necessary parameters to construct 
$L^2$-orthogonal STC. The modulation index $h$ and the phase smoothing
function $q(t)$ can be chosen with the usual restrictions of
conventional CPM detailed above.

As proved by our simulations in section \ref{sec:sim}, the values of
the initial phases $\theta_m(1)$, which are known to have no influence
in conventional CPM systems \cite{Ande86}, will be shown to have instead a
great importance on the performance of the proposed code.

\subsection{Continuous-time Model}

First, we need to introduce another formalism than the block structure
used for design. The signals sent by each Tx antenna $m$ are
rewritten as
\begin{equation}
  \underline
  s_m(t)= \sqrt{\frac{E_s}{L_tT}} \exp(j2\pi[\theta_m(1) + 
  h\sum\limits_{i=1}^{N_c}d_i q(t-(i-1)T) +c_m(t)]).
  \label{eq:smcont}
\end{equation}

\begin{figure}
  \centering
  \includegraphics[width=0.5\columnwidth]{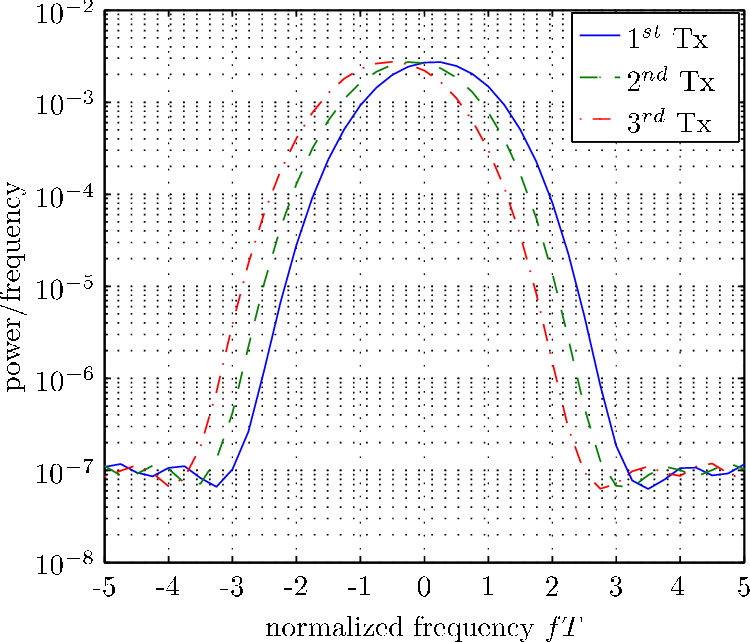}
  \caption{Power spectral density of the linPC analyzed with the Welch
    algorithm}
  \label{fig:PSS}
\end{figure}
In Eq. (\ref{eq:smcont}) the phase memory terms $\theta_m(L_tl+r)$ get
included in the summation term.  Only the initial phase $\theta_m(1)$
remains. This is due to the property of continuity in the definition
of every symbol over the whole time $N_cT$ where $N_c$ is the number
of transmitted symbols. As a result of the parallel mapping, the data
symbols $d_i$ are also equal for each antenna. For offPC codes, we
use the modified alphabets detailed in section \ref{sec:ortho} and no
additional correction factor is necessary. For linPC codes, the
correction factor $c_m(t)$ simplifies to a continuous linear function
$c_m(t)=(m-1)t/L_t$.

It is interesting to see that for the second Tx antenna, the correction
factor causes a constant phase offset of $2\pi/L_t$ per symbol and of
$2\pi$ per block. For the 3 Tx antennas case, these offsets are
multiplied by 2. The same effect is observed from the offset added to
the alphabet $\Omega_{d_m}$.

This phase offset induces a frequency shift. For a phase shift of
$2\pi(m-1)/L_t$ on a period of $T$, we get a frequency shift of
$\Delta f_m=(m-1)/TL_t$ for antenna $m$ and a symbol length $T$.

Figure \ref{fig:PSS} shows the simulated power spectral density for
the linPC code with 3 Tx antennas, $h=1/2$, $g=2$ and $M=4$. To
achieve an attenuation in power of -30dB, a bandwidth expansion of
some $5\,\mbox{Hz} T $ is necessary. This corresponds to an increased
bandwidth demand of $\frac{\Delta f_2
  T}{5\mbox{Hz}T}=\frac{2}{3\cdot5}=0.133$. As the alphabet size $M$
grows, the absolute bandwidth of the CPM signal widens but the shift
caused by the coding scheme is constant. Therewith the additional
relative bandwidth eventually decreases.

It should be recalled here that the highest achievable rate for linear
codes with 4 Tx antennas is 3/4 \cite{Taro99a}. It means that to
transmit the same quantity of data during the same duration of time as
for our proposed code a 25\% increase in bandwidth is necessary.

With this continuous-time formalism, the signals sent can be rewritten
in vector form as $\underline s(t)=[\underline s_1(t),\underline
s_2(t),\cdots,\underline s_{L_t}(t)]^{\sfT}$. Below underlined
variables denote the vector representation and non-underlined
variables the previously used matrix form. Thus we can write
$\underline s(t)$ as the product of two matrices $\underline\Theta$
and $\underline C(t)$ and a vector $\underline d(t)$
\begin{equation}
  \underline s(t)=\sqrt{\frac{E_s}{L_tT}}~ \underline\Theta\,\underline C(t) \,\underline d(t)
  \label{eq:sVec}
\end{equation}
The matrix of initial values $\underline
\Theta=\diag(\underline\theta)$ and the matrix of correction factors
$\underline C(t)=\diag(\underline c(t))$ are $L_t\times L_t$ diagonal
matrices obtained from the vectors
\begin{align}
  \underline\theta=&\begin{bmatrix}
    \exp(j2\pi\theta_1(1))\\
    \exp(j2\pi\theta_2(1))\\
    \vdots\\
    \exp(j2\pi\theta_{L_t}(1))\\
  \end{bmatrix},& \hspace*{-0.5em}\underline c(t)=&\begin{bmatrix}
    \exp(j2\pi c_1(t))\\
    \exp(j2\pi c_2(1))\\
    \vdots\\
    \exp(j2\pi c_{L_t}(1))\\
  \end{bmatrix}.
\end{align}
As a result of the parallel mapping, the vector of data symbols can be
written as
\begin{equation}
  \underline d(t)=\exp(j2\pi h\sum\limits_{i=1}^{N_c} d_iq(t-(i-1)T))
  {\begin{bmatrix} 1 & 1 & \hdots& 1
    \end{bmatrix}}^{\sfH}.
\end{equation}

\subsection{Code Performance}

The performance of theses codes may now by evaluated using the
classical pair-wise error probability (PWEP). We assume optimal
demodulation, i.e.  maximum likelihood (ML) sequence detection (MLSD).
Furthermore, it is considered that for $0\leq t\leq N_cT$ the signal
$\underline s(t)$ modulated by the data sequence $d_j$ is the one truly
sent.  The PWEP is then the probability that this signal is
erroneously detected as signal $\hat{\underline s}(t)$ modulated by $\hat
d_j$ \cite{Zaji05}.
\begin{equation}
  PWEP=P(\underline s(t)\rightarrow\hat{\underline s}(t)|\mb A)=Q\left(\frac{\| \mb A^{\sfT}\mb\Delta(t)\|}{\sqrt{2N_0}}\right)
\end{equation}
where $\mb A$ is the channel matrix which is assumed to to have
frequency flat quasi-static Rayleigh fading and mutual independent
elements. The energy of noise is given by $N_0$ and $Q(\cdot)$ is the
cumulative distribution function of the normal distribution
(Q-function). The normalized difference vector $\mb\Delta(t)$ is given by
\begin{equation}
  \mb\Delta(t)=\begin{bmatrix}
    \Delta_1(t)\\ \Delta_2(t)\\ \vdots \\ \Delta_{L_t}(t)
  \end{bmatrix}= \sqrt{\frac{L_tT}{E_s}}
  \begin{bmatrix}
    \underline s_1(t)-\hat{\underline s}_1(t)\\ \underline s_2(t)-\hat{\underline s}_2(t)\\ \vdots \\
    \underline s_{L_t}(t)-\hat{\underline s}_{L_t}(t)
  \end{bmatrix}
  \label{eq:Delta}
\end{equation}

The PWEP is minimized by maximizing the product of the eigenvalues of
the signal matrix \cite{Zhan03,Zaji05}

\begin{equation}
  \mb C_s=\left[\begin{smallmatrix}
      \int\limits_0^{N_cT}|\Delta_1(t)|^2\D t& \int\limits_0^{N_cT}\hspace{-0.5em}\Delta_1(t)\Delta_2^*(t)\D t&\cdots& \int\limits_0^{N_cT}\hspace{-0.5em}\Delta_1(t)\Delta_{L_t}^*(t)\D t \\
      \int\limits_0^{N_cT}\hspace{-0.5em}\Delta_2(t)\Delta_1^*(t)\D t& \int\limits_0^{N_cT}|\Delta_2(t)|^2\D t &\cdots& \int\limits_0^{N_cT}\hspace{-0.5em}\Delta_2(t)\Delta_{L_t}^*(t)\D t \\
      \vdots&&&\vdots\\
      \int\limits_0^{N_cT}\hspace{-0.5em}\Delta_{L_t}(t)\Delta_1^*(t)\D t& \int\limits_0^{N_cT}\hspace{-0.5em}\Delta_{L_t}(t)\Delta_2^*(t)\D t&\cdots& \int\limits_0^{N_cT}|\Delta_{L_t}(t)|^2\D t \\
    \end{smallmatrix}\right]
  \label{eq:cs}
\end{equation} 
Using Eq. (\ref{eq:sVec}) and (\ref{eq:Delta}) the signal matrix can
be written as
\begin{figure*}
  \subfigure[offset PC with
  REC]{\includegraphics[width=5.9cm]{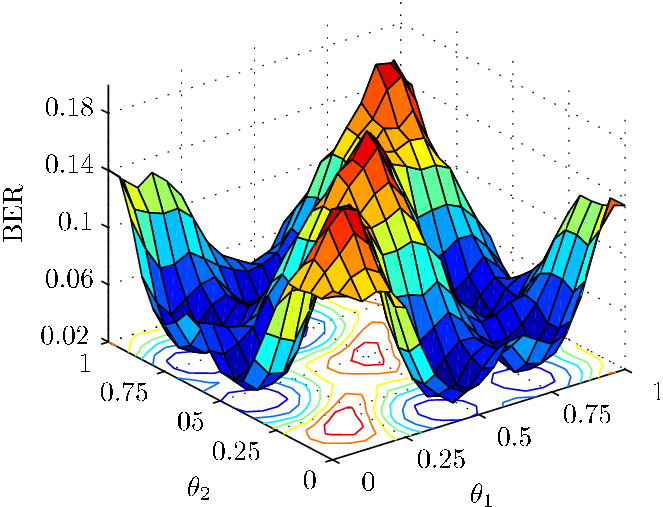}\label{fig:ab1}}
  \subfigure[offset PC with
  RC]{\includegraphics[width=5.9cm]{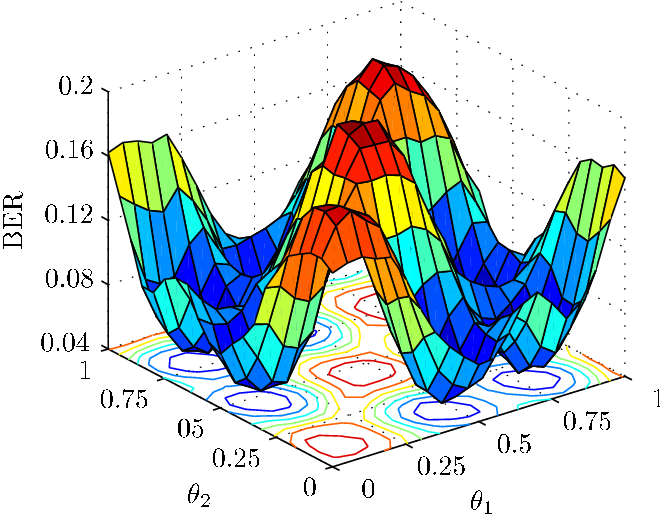}\label{fig:ab2}}
  \subfigure[linear PC with
  REC]{\includegraphics[width=5.9cm]{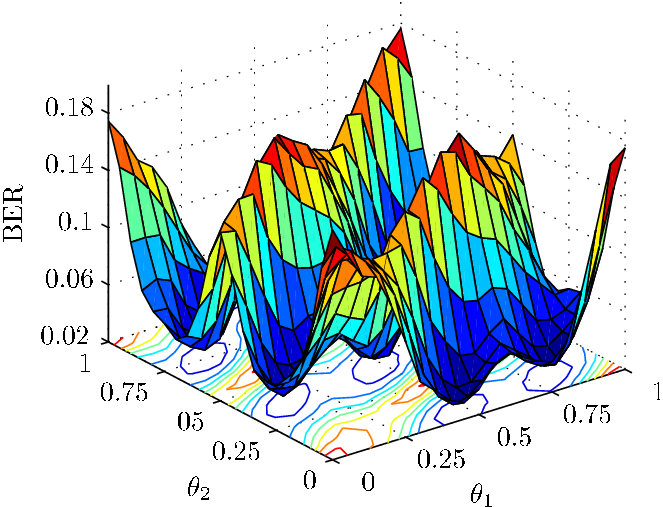}\label{fig:ab3}}
  \caption{Simulated BER for varying initial phase $\theta_1(1)$ and
    $\theta_2(1)$ and $\theta_3(1)=0$ for 3 Tx antennas at 13dB SNR}
  \label{fig:ab}
\end{figure*}
\begin{align}
  \mb C_s=&\int\limits_0^{N_cT}\!\mb\Delta(t)\mb\Delta^{\sfH}(t)\D t
  \nn =&\int\limits_0^{N_cT}\!{\big[\underline\Theta\,\underline
    C(t)(\underline d(t)-\hat{\underline d}(t))\big]}
  \big[\underline\Theta\,\underline C(t)(\underline d(t)-\hat{\underline
    d}(t))\big]^{\sfH} \D t\nn
  =&~\underline\Theta\big(\!\!\int\limits_0^{N_cT}\!\underline
  C(t)\,\Delta\underline d(t)\,\Delta \underline
  d^{\sfH}(t)\,\underline C^{\sfH}(t) \D t
  \big)\underline\Theta^{\sfH}
\end{align}
with $\Delta\underline d(t)=\underline d(t)-\hat{\underline d}(t)$ and
the integral over the matrix acting element-wise. Since $\underline
d(t)$ has equal elements and is multiplied by its Hermitian transpose,
we get an $L_t\times L_t$ all-ones matrix. This matrix is multiplied
with the matrices of the correction factor and we obtain the
correlation matrix of the correction vector
\begin{equation}
  \mb C_s=\underline\Theta\!\int\limits_0^{N_cT}\!\underline c(t)\underline c^{\sfH}(t)\D t~ \underline\Theta^{\sfH}.
\end{equation}
By writing $c_m(t)=\frac{m-1}{L_r}\bar c(t)$ in Eq. (\ref{eq:c1}) and
(\ref{eq:c2}), we get
\begin{equation}
  \underline c_m(t)\underline c^*_{m'}(t)=\exp\big(j2\pi\bar
  c(t)({\textstyle \frac{m-1}{L_t}-\frac{m'-1}{L_t}})\big).
\end{equation}
The autocorrelation $\underline c_m(t)\underline c_m^*(t)$ is therewith
always one and we have
\begin{align}
  &\underline c(t)\underline c^{\sfH}(t)=
  \left[\begin{smallmatrix}
      \exp\!\big(\frac{0}{L_t}j2\pi\bar c(t)\big)&\exp\!\big(-\frac{1}{L_t}j2\pi\bar c(t)\big)&\hdots &\exp\!\big(-\frac{L_t-1}{L_t}j2\pi\bar c(t)\big)\!\\
      \exp\!\big(\frac{1}{L_t}j2\pi\bar c(t)\big)&\exp\!\big(\frac{0}{L_t}j2\pi\bar c(t)\big)&\hdots &\exp\!\big(-\frac{L_t-2}{L_t}j2\pi\bar c(t)\big)\!\\
      \vdots&&&\vdots\\
      \!\exp\!\big(\frac{L_t-1}{L_t}j2\pi\bar c(t)\big) &\exp\!\big(\frac{L_t-2}{L_t}j2\pi\bar c(t)\big)&\hdots &\exp\!\big(\frac{0}{L_t}j2\pi\bar c(t)\big)\\
    \end{smallmatrix}\right]
\end{align}
The elementwise integration of this matrix is easily computed \cite{Hess08g}
in the special cases where $\bar c(t)$ is a linear function (linPC
codes) or a sum of raised cosines (OffPC codes) \cite{Hess08g}. In
both cases,  the matrix $ \mb C_s$ is shown to have full rank \cite{Hess08} and
thus our codes achieve full diversity. However, the PWEP approach
doesn't provide here any valuable estimate of the coding gain. For that
reason, we detail hereafter some statistical estimates
to show the influence of the initial phase values upon the coding
gain.

\section{Simulations}
\label{sec:sim}

In this section, we benchmark by simulations the performance of the
proposed codes.  In all the simulations, we used an alphabet size of
$M=4$ with 2-bit Gray-coding, a modulation index of $h=1/2$, a memory
length of $\gamma=2$ and 12 samples per symbol. The signal is
disturbed by complex code-block-wise Rayleigh fading of variance one.
In this section all given phase values are relative values, e.g. 1
corresponds to $2\pi$ or $360^\circ$.

\subsection{Two transmit antennas}

\begin{figure}
  \centering
  \includegraphics[width=0.5\columnwidth]{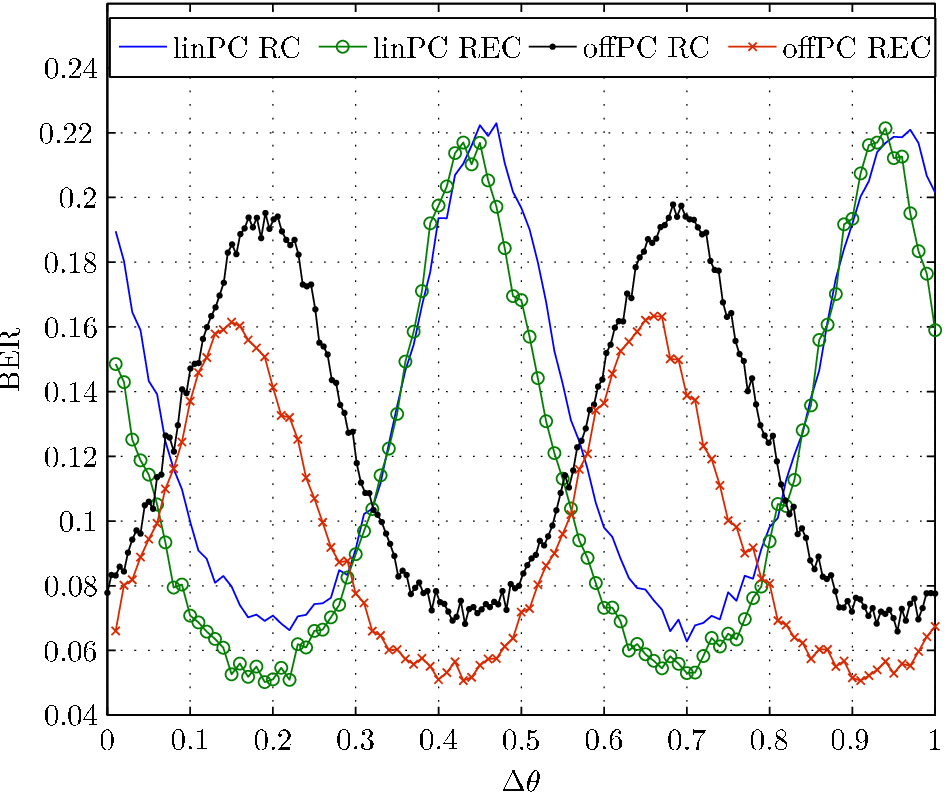}
  \caption{simulated BER for varying initial phase $\theta_2$ for 2 Tx
    antennas at SNR=12.5dB }
  \label{fig:a}
\end{figure}

The bit error rate (BER - $E_b/N_0$) for the proposed two transmitting
antenna codes depends on the difference of the initial phase
$\Delta\theta=\theta_2(1)-\theta_1(1)$. Figure \ref{fig:a} shows the
results of computer simulations for linPC and offPC codes with
different phase smoothing functions $q(t)$. The variation of
performance covers almost one decade. This shows the importance of a
carefully chosen initial phase.

Mainly, the position of the minimal BER seems to depend on the
correction factor used. Between linPC and offPC, the minima are
shifted by $1/4$. However, the phase smoothing function used has only
minor influence on the position of the minima. It is also interesting
to see that the distance between the minima is $1/2$ and further
simulations show a periodicity of $1$.

\subsection{Three transmit antennas}

\begin{table}\normalsize\centering
  \begin{tabular}{cc|cccccc}
    \multirow{2}{*}{offPC}&$\theta_1$&0.1&0.15&0.4&0.45&0.75&0.8\\
    &$\theta_2$&0.45&0.75&0.8&0.1&0.15&0.4\\ \hline
    \multirow{2}{*}{linPC}&$\theta_1$&0.75&0.4&0.45&0.7&0.05&0.1\\ &$\theta_2$&0.15&0.15&0.5&0.8&0.5&0.8\\ 
  \end{tabular}
  \caption{Initial phase values for minimal BER (SNR=12.5dB; 2REC)}\vspace{-2em}
  \label{tab:min}
\end{table}

For the three antennas codes, the BER also depends on the initial phase
$\theta_m(1)$.  Fig. \ref{fig:ab} shows the simulation results for
different codes with varying initial phases for the first and second
antenna and null initial phase for the third antenna. It can be seen
that the phase offset for a minimal BER depends on the correction
factor chosen (Fig. \ref{fig:ab1} and \ref{fig:ab3}). However,
similarly to the two antenna code, the form of the phase smoothing
function $q(t)$ has almost no influence on the position of the minima
(Fig. \ref{fig:ab1} and \ref{fig:ab2}). In Table \ref{tab:min} the
optimal initial phase are summarized for the first and second antenna.
Other simulations for different alphabet sizes $M$, modulation indexes
$h$ and memory length $\gamma$ validate the position of the minima and
prove it is an important issue for the optimal design of parallel
codes.

\subsection{Bit Error Rate}

\begin{figure}
  \centering
  \includegraphics[width=0.5\columnwidth]{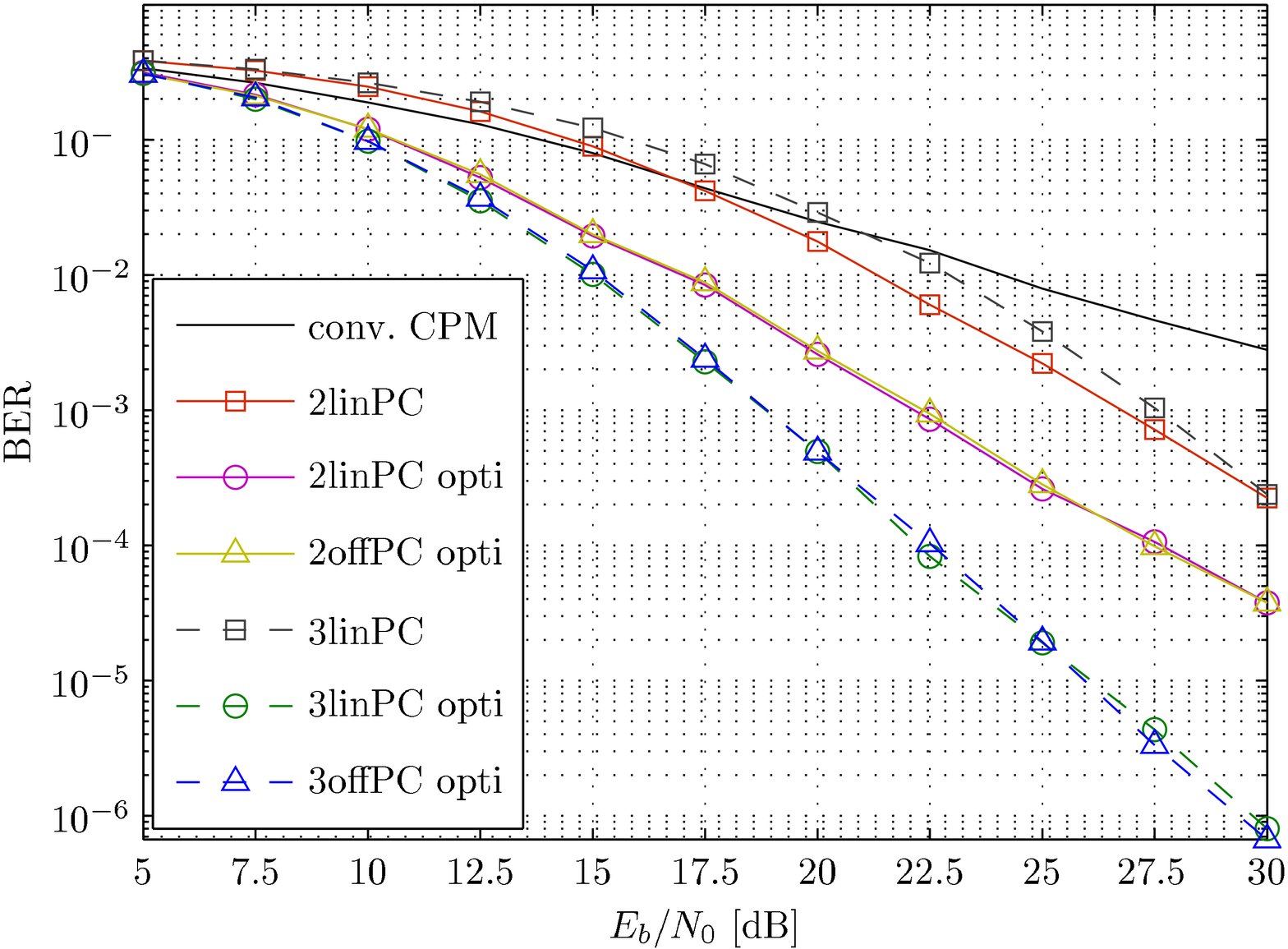}
  \caption{BER for Rayleigh-fading channel}
  \label{fig:BER}
%  \vspace{-1.6em}
\end{figure}

Fig. \ref{fig:BER} shows the influence of Rayleigh fading channels on
a CPM transmitter with a different number of antennas.

For the optimal two antenna system we used in the optimal case a
frequency offset of $\Delta\theta=0.19$ for linPC and of
$\Delta\theta=0.4$ for offPC. For the optimal three-antennas-codes, we
took the following values from Fig. \ref{fig:ab}:
\begin{itemize}
\item offPC: $\theta_1(1)=0.1$, $\theta_2(1)=0.45$;
\item linPC: $\theta_1(1)=0.4$, $\theta_2(1)=0.15$.
\end{itemize}
The non-optimal codes have no phase offset
($\theta_1(1)=\theta_2(1)=\Delta\theta=0$).

The optimized codes achieve the expected performance gain.  For high
SNR the BER decreases with 5dB/dec similar to a two antenna system
with full diversity. The three Tx antennas code achieves a decay of
some 3.5dB/dec. This validates the property of full diversity.

Fig. \ref{fig:BER} shows clearly the improvement of the coding gain by
using an optimized initial phase. Comparing the optimized codes with
the non-optimal ones, we achieve an additional coding gain of around
5dB for the two antenna system and of around 7dB for the three Tx
antennas.

% Also the systematical character of the minimal BER and the
% associated geometric pattern can be analytically limited to the
% maximization of the coding gain.
% 
% % we expect a closed analytical solution which may result out of the
% % conditions associated with the coding gain. For the two antenna
% % systems this effect is not observed.
% 
% Fig. \ref{fig:BER} shows the influence of additive white Gaussian
% noise. For the first two three antenna codes we use the minimal
% values from Fig. \ref{fig:ab} (offPC: $\theta_1(t)=0.45$,
% $\theta_2(t)=0.1$; linPC: $\theta_1(t)=0.15$, $\theta_2(t)=0.75$)
% and for the third three antenna code we us a not optimized PC
% ($\theta_1(0)=\theta_2(0)=0$). As well for the two antenna code as
% for all three antenna codes the gain through diversity and
% optimization is obvious. By means of the slope of the BER we see
% that as well the optimized as the not optimized codes reach full
% diversity. The increased performance of the optimized code relates
% to an improved coding gain.
% 
% \begin{figure}
%   \centering
%   \includegraphics[width=6.5cm]{BER12333.pdf}
%   \caption{BER for Rayleigh-fading and AWGN}
%   \label{fig:BER}
% \end{figure}
% 
% % \subsection{Frequency shift}
% 
% The correction factor causes a frequency shift due to a
% non-zero-mean value. But in \cite{Hess08} we that this shift is
% quite moderate. Here we will compare the simulated and the
% calculated frequency shift.
	
\section{Conclusion}

In this paper, we detail the construction and analyze some of the
properties of L2-orthogonal STC-CPM for two and three transmitting
antennas.  These codes are attractive due to their low-effort-decoding
and the few restrictions the code-family set upon the parameters of
CPM.  We give a general formulation for two and three antenna parallel
codes and introduce a continuous-time representation of the CPM
signals.  With this representation we are able to  analyze how  the coding
gain depends on the initial phase of the system. Furthermore, we give
the optimal values for the initial states obtained from computer
simulation. The significant gain in performance for typical Rayleigh
fading channels is shown and compared with non-optimal parallel codes.

\bibliographystyle{IEEEtran} \bibliography{IEEEabrv,bib}

\end{document}